# Observation of pseudospin-locked chiral edge transport in gyromagnetic photonic crystals

Chengle Fan[1†], Kaizhou Shi[1†], Xiang Xi[2,*], Zhen Gao[1,*]

[1]State Key Laboratory of Optical Fiber and Cable Manufacturing Technology, Department of Electronic and Electrical Engineering, Guangdong Key Laboratory of Integrated Optoelectronics Intellisense, Southern University of Science and Technology, Shenzhen 518055, China.

[2]School of Electrical Engineering and Intelligentization, Dongguan University of Technology, Dongguan, 523808, China.

[†]These authors contributed equally: Chengle Fan and Kaizhou Shi

*Corresponding author(s). E-mail(s): xix@dgut.edu.cn; gaoz@sustech.edu.cn;

**Photonic pseudospins enable selective excitation and routing of light, but perturbations that mix opposite pseudospins can induce backscattering in helical edge channels. Here, we experimentally demonstrate pseudospin-locked chiral edge transport in gyromagnetic photonic crystals, combining pseudospin selectivity with nonreciprocal unidirectional propagation. By jointly engineering lattice deformation and magnetic bias, we control dipolar–quadrupolar band inversions and construct interfaces supporting pseudospin-selective chiral channels. Microwave near-field and transmission measurements reveal a single chiral edge branch that is selectively excited by only one source handedness and exhibits pronounced nonreciprocal transport. Extending this mechanism to multidomain structures, we combine two pseudospin channels into a common waveguide, separate them into distinct outputs, and route them through intersecting interfaces. These configurations realize a nonreciprocal photonic pseudospin multiplexer and demultiplexer, together with a waveguide crossing exhibiting average measured crosstalk of −14.1 and −15.8 dB for its two input channels. Our results present an experimental route to integrating pseudospin-selective channel control with chiral edge transport, providing a foundation for nonreciprocal photonic networks that use pseudospin to organize and route multiple wave channels.**

## Introduction

Photonic networks require reliable wave transport together with a controllable degree of freedom for distinguishing, combining, and routing multiple channels. Topological photonics offers a route towards this goal by linking interface modes to the topology of bulk bands. Among the available degrees of freedom, photonic pseudospin provides a synthetic mode label that enables selective excitation and directional control. Pseudospin-dependent edge transport has been explored in metamaterials, metawaveguides, coupled-resonator arrays and dielectric photonic crystals [1-8]. In hexagonal photonic crystals, combinations of dipolar and quadrupolar modes with opposite phase winding define two pseudospin sectors. Inverting the dipolar − quadrupolar band ordering produces a photonic analogue of the spin Hall phase, whose helical edge channels associate opposite pseudospins with opposite propagation directions [7,8]. This connection allows the handedness of a local excitation to control the direction of guided waves, making pseudospin an attractive degree of freedom for channel-selective photonic transport.

The robustness of such helical channels (Fig. 1a,d), however, depends on the symmetries and modal structure that distinguish the two pseudospins. Perturbations that couple opposite pseudospin sectors can scatter waves between counterpropagating edge modes, compromising directional transport [7-9]. Physical time-reversal symmetry alone therefore does not guarantee the protection of these synthetic photonic degrees of freedom [9,10]. Breaking time-reversal symmetry offers a complementary route: photonic Chern phases support chiral edge channels with nonreciprocal propagation (Fig. 1b,e) [11-16]. At an interface supporting a single chiral branch within a common bulk gap, the absence of a counterpropagating edge partner suppresses backscattering. The Chern number constrains the net chirality of the interface spectrum, while assigning a controllable pseudospin character to its channels requires additional modal engineering. Combining these capabilities would allow pseudospin to organize multiple channels while their propagation remains chiral and nonreciprocal (Fig. 1c,f).

Recent studies have demonstrated the value of combining chiral transport with an additional wave degree of freedom. One-way valley-polarized transport and chiral valley edge states have enabled valley-selective propagation, multiplexing, and interface crossings [17,18]. Co-propagating topological interface modes have also been demonstrated through engineered pseudospin-flipping coupling in microring arrays, enabling gain-controlled power transfer and switching [19]. These developments motivate a complementary approach based on the dipolar–quadrupolar pseudospins of photonic crystals. Whereas valley labels distinguish inequivalent regions of momentum space, these pseudospins are defined by phase-winding combinations of lattice modes near the Brillouin-zone centre. Theoretical work has predicted that geometric deformation and magnetic bias can produce pseudospin-polarized topological phases and edge channels in gyromagnetic photonic crystals [20,21]. However, experimental observation of pseudospin-locked chiral edge transport in gyromagnetic photonic crystals remains elusive, and its extension to pseudospin-selective routing through junctions and intersecting interfaces has yet to be demonstrated.

Here, we experimentally observe pseudospin-locked chiral edge transport in gyromagnetic photonic crystals. By jointly engineering lattice deformation and magnetic bias, we control dipolar–quadrupolar band inversions in the two pseudospin sectors and construct interfaces supporting pseudospin-selective chiral channels. Microwave near-field measurements resolve a single chiral edge branch selectively excited by one source handedness, while transmission measurements show pronounced nonreciprocal propagation. Together, these observations connect the pseudospin character of the lattice modes, their selective coupling to local excitations, and the propagation direction of an unpaired chiral edge channel. We further translate this mechanism into multidomain photonic structures that combine, separate, and route distinct pseudospin channels. A nonreciprocal pseudospin multiplexer combines two input channels into a common waveguide, while a complementary demultiplexer directs opposite pseudospins towards distinct outputs. An interface crossing routes the two channels through an intersection, with average measured crosstalk of −14.1 and −15.8 dB for its two inputs. These demonstrations extend pseudospin-selective chiral transport from an isolated interface to interconnected routing elements. Our results establish an experimental connection between pseudospin-resolved band engineering and nonreciprocal network operation, providing a foundation for photonic networks in which pseudospin serves as an addressable label for organizing multiple wave channels.

## Results

### Topological phase engineering in a gyromagnetic photonic crystal

To realize pseudospin-locked chiral edge states, we consider a two-dimensional (2D) gyromagnetic photonic crystal comprising hexagonal six-rod clusters arranged on a triangular Bravais lattice with a lattice constant of $a$ = 14 mm. Each unit cell contains six yttrium iron garnet (YIG) rods with a diameter of $d$ = 2.8 mm, each sandwiched between two permanent magnets, as illustrated in Fig. 2a. The unit-cell deformation is characterized by the ratio $a/R$, where $R$ is the distance from the center of each rod to the unit-cell center. By jointly varying the external bias magnetic field $H_0$ and the deformation parameter $a/R$, we map the topological phases of the crystal. The magnetic bias induces gyromagnetic anisotropy in the YIG rods and breaks time-reversal symmetry, while the geometric deformation modifies the relative ordering of the dipolar and quadrupolar modes. The resulting phase diagram (Fig. 2b) reveals three distinct phases: photonic quantum spin Hall (QSH), pseudospin-polarized quantum anomalous Hall (PQAH), and conventional insulator (CI).

We next fix the external bias magnetic field at $H_0$ = 2200 Oe and vary only $a/R$, following the horizontal dashed line in Fig. 2b. Figure 2c shows the evolution of the four relevant transverse magnetic (TM) eigenfrequencies at the Γ point. Two successive band inversions, each accompanied by gap closure and reopening at Γ, divide the parameter range into three distinct topological regimes. To elucidate their orbital character and topology, we examine representative bulk band structures in the QSH, PQAH, and CI phases, shown in Fig. 2d–f, respectively. We first consider the PQAH structure at $R$ = $a$/3, which exhibits a complete photonic bandgap (Fig. 2e). The corresponding $E_z$ field and phase distributions of the Γ-point eigenstates are shown beneath the band structure. The spatial symmetries of the $E_z$ field can be described using the dipolar orbital basis functions $p_x$ and $p_y$ and the quadrupolar orbital basis functions $d_{xy}$ and $d_{x^2-y^2}$. Their complex combinations, $p_\pm = p_x \pm jp_y$ and $d_\pm = d_{x^2-y^2} \pm jd_{xy}$, where $j^2 = -1$, define the pseudospin basis $[p_+, d_+, p_-, d_-]$. Here, $p_+$ and $d_+$ represent pseudospin-up states, whereas $p_-$ and $d_-$ represent pseudospin-down states. The two pseudospins are distinguished by their opposite phase winding: the phase increases anticlockwise for pseudospin-up states and clockwise for pseudospin-down states, accumulating $2\pi$ for the dipolar modes and $4\pi$ for the quadrupolar modes around the unit-cell center.

We characterize the topology of the magnetically biased system using the spin Chern numbers $(C_+, C_-)$, which quantify the contributions of the two pseudospin sectors [20]. For the PQAH structure at $R = a/3$, the Γ-point eigenstates are ordered as $[p_+, d_-, p_-, d_+]$ in ascending eigenfrequency, and the corresponding spin Chern numbers are $(C_+, C_-) = (0, -1)$. Increasing $R$ from this configuration drives a topological transition at $R = a/2.83$, where the $p_+$ and $d_+$ states undergo band inversion. In the resulting QSH phase, the spin Chern numbers are $(C_+, C_-) = (1, -1)$. Figure 2d shows the band structure and the corresponding $E_z$ field and phase distributions for a representative QSH structure at $R = a/2.7$, where the eigenstates are ordered as $[d_-, d_+, p_+, p_-]$.

Conversely, decreasing $R$ from the PQAH configuration induces a band inversion between the $p_-$ and $d_-$ states, driving the system into the CI phase. Figure 2f shows the band structure and the corresponding $E_z$ field and phase distributions at $R = a/3.5$. The eigenstates are ordered as $[p_+, p_-, d_-, d_+]$, with spin Chern numbers $(C_+, C_-) = (0, 0)$, consistent with a topologically trivial phase. All eigenstate orderings are given in ascending eigenfrequency at Γ. Reversing the bias magnetic field leaves the band structure unchanged but interchanges the two pseudospin sectors and reverses the signs of their Chern numbers, such that $(C_+, C_-) \rightarrow (-C_-, -C_+)$. The PQAH phase therefore has $(C_+, C_-) = (1, 0)$ under the reversed bias $(\otimes B)$, whereas the QSH and CI phases retain their respective topological classifications (Fig. 2b).

**Experimental observation of pseudospin-locked chiral edge states**

We now experimentally demonstrate the pseudospin-locked chiral edge state. According to bulk–edge correspondence, the difference in total Chern number across an interface determines the net chirality of its edge states. When the two pseudospin sectors remain effectively decoupled, the spin Chern number differences $\Delta C_\pm$ further resolve the contributions from each sector, providing a framework for predicting pseudospin-locked chiral edge states. Here, we consider an interface between PQAH and CI domains with $R = a/3$ and $R = a/4.05$, respectively (Fig. 3a). The CI structure used here has the same spin Chern numbers as the structure at $a/R = 3.50$ in Fig. 2f, despite a reversal of the relative frequency ordering of the $p_+$ and $p_-$ states below the bandgap of interest. Defining $\Delta C_\pm = C_\pm^{\mathrm{PQAH}} - C_\pm^{\mathrm{CI}}$, we obtain $(\Delta C_+, \Delta C_-) = (0, -1) - (0, 0) = (0, -1)$, predicting a single chiral edge state in the pseudospin-down sector. Consistent with this prediction, the calculated projected band structure exhibits a single chiral branch within the common bulk bandgap (Fig. 3b). We experimentally demonstrate the pseudospin-selective excitation of this chiral edge state below.

Figure 3a shows the photograph of the fabricated experimental sample, with the PQAH domain below the interface and the CI domain above it. Their unit cells are highlighted in blue and purple, respectively, and the inset shows the YIG rod sandwiched between two permanent magnets. We first measure the transmission spectra by placing source and probe antennas at the two ends of the interface (Fig. 3c), where $S_{21}$ and $S_{12}$ denote rightward and leftward transmissions, respectively. Within the common bulk bandgap (grey region), the rightward transmission (blue line) is approximately 20 dB higher than the leftward transmission (red line), demonstrating nonreciprocal transport of the chiral edge states along the $+x$ direction.

To demonstrate the pseudospin selectivity of the chiral edge state, we replace the point-source antenna with a chiral-source antenna that generates clockwise or anticlockwise phase vortices (Fig. 3d), enabling selective excitation of the two photonic pseudospins. We measure the complex $E_z$ field distributions for each source chirality and retrieve the corresponding dispersion relations by spatial Fourier transformation (Fig. 3e–h). A pseudospin-down source ($S_-$) excites the right-propagating chiral edge state (Fig. 3f,h), whereas a pseudospin-up source ($S_+$) produces no observable guided edge-state excitation (Fig. 3e,g). These observations agree with the simulations and confirm both pseudospin locking and unidirectional transport of the chiral edge state.

To test the robustness of the pseudospin-polarized chiral edge states, we replace a subset of the YIG rods along the interface with aluminum pillars and repeat the near-field measurement using the same pseudospin-down excitation as in Fig. 3h. The rightward-propagating chiral edge state persists in the presence of these metallic defects (Fig. 3i), demonstrating its robustness against strong perturbations.

**Experimental demonstration of nonreciprocal pseudospin-selective topological devices**

Having experimentally established the pseudospin-locked chiral edge state at the PQAH–CI interface, we next use the independently controllable spin Chern numbers to construct pseudospin-selective photonic routing devices. By spatially arranging domains with different $(C_+, C_-)$, we combine and separate the two pseudospin components and route them through a waveguide crossing with low crosstalk.

We first construct a nonreciprocal photonic pseudospin multiplexer comprising three domains with $(C_+, C_-) = (1, -1)$, $(1, 0)$, and $(0, -1)$ (Fig. 4a). The interfaces connected to ports P1 and P2 have spin Chern number differences of $(\Delta C_+, \Delta C_-) = (0, -1)$ and $(1, 0)$, respectively, supporting a pseudospin-down input channel at P1 and a pseudospin-up input channel at P2. The output interface connected to P3 has $(\Delta C_+, \Delta C_-) = (1, 1)$, allowing both pseudospin components to propagate in the same direction along a common output channel. The simulated field distribution under simultaneous excitation at P1 and P2 shows that the two input waves combine and propagate toward P3 (Fig. 4b). The simulated $|E_z|$ distributions under separate excitation at P1 and P2 likewise show rightward propagation toward P3 (Figs. 4c,e), while the phase winding in the output channel confirms that each wave retains its pseudospin character. Reverse-excitation simulations, with a source of the corresponding chirality placed at P3, show strongly suppressed leftward propagation (Fig. 4d,f).

To characterize the multiplexer experimentally, we excite P1 and P2 separately and measure the transmission to the other two ports (Fig. 4g,h). For the pseudospin-down input at P1, $S_{31}$ describes transmission to the common output P3, whereas $S_{21}$ measures leakage into P2. Within the operating band, the measured $S_{31}$ (red solid line) exceeds $S_{21}$ (blue solid line) by an average of 26 dB, confirming preferential transmission to P3 (Fig. 4g). For the pseudospin-up input at P2, the measured $S_{32}$ (red solid line) exceeds $S_{12}$ (blue solid line) by an average of 28 dB, demonstrating transmission to P3 with suppressed leakage into P1 (Fig. 4h). These measurements agree with the simulated spectra (dashed lines) and demonstrate pseudospin multiplexing into a common output channel with low interchannel crosstalk.

A complementary spatial arrangement of the same three topological domains forms a photonic pseudospin demultiplexer (Fig. 4i). The common input interface connected to P1 supports both pseudospin components, whereas the two output interfaces selectively support opposite pseudospins. The pseudospin-up component is therefore routed to P2, and the pseudospin-down component to P3, as illustrated by the simulated field distribution in Fig. 4j. The simulated $|E_z|$ distributions under separate excitation confirm that a pseudospin-up source excites the P1–P2 channel (Fig. 4k), whereas a pseudospin-down source excites the P1–P3 channel (Fig. 4m). Under reverse excitation at P2 or P3, the simulated field remains concentrated near the driven port, with strongly suppressed propagation toward P1 (Fig. 4l,n).

We experimentally characterize this pseudospin-selective routing by exciting the common input P1 with opposite source chiralities and measuring $S_{21}$ and $S_{31}$, which describe transmission to P2 and P3, respectively (Fig. 4o,p). Under pseudospin-up excitation, the measured $S_{21}$ (blue solid line) exceeds $S_{31}$ (red solid line) by an average of 20 dB within the operating band, indicating preferential transmission to P2 with suppressed leakage into P3 (Fig. 4o). Under pseudospin-down excitation, this transmission contrast is reversed: $S_{31}$ exceeds $S_{21}$ by an average of 26 dB, indicating preferential transmission to P3 (Fig. 4p). The measured spectra agree with the simulations (dashed lines), demonstrating a photonic pseudospin demultiplexer that directs opposite input pseudospins to distinct output ports. Together, these measurements establish the multiplexing and demultiplexing of photonic pseudospins through topologically defined routing channels.

We further realize a pseudospin-locked waveguide crossing by combining four domains with $(C_+, C_-) = (1,0)$, $(1,-1)$, $(0,0)$, and $(0,-1)$ (Fig. 5a). The horizontal channel from P1 to P3 supports only the pseudospin-up edge states, whereas the intersecting channel from P2 to P4 supports only the pseudospin-down edge states. At the intersection, the unintended output arm provides no in-gap channel with the same pseudospin as the incident wave. Coupling into this arm therefore requires pseudospin conversion and thus is strongly suppressed. The simulated field distribution under simultaneous excitation shows that the two waves traverse the intersection along their respective channels (Fig. 5b). Under separate excitation at P1 and P2, the simulated and measured field distributions at 11.00 GHz confirm transmission toward P3 and P4, respectively, with little leakage into the other channel (Fig. 5c,d and 5g,h).

The measured transmission spectra further confirm low-crosstalk operation across the operating band (Figs. 5e,f and 5i,j). For each input port, we average each transmission spectrum in dB over 11.0–11.5 GHz. We then define the crosstalk as the larger of the two mean transmissions at the unintended receiving ports minus the mean transmission at the intended output. The resulting simulated and measured crosstalk levels are approximately −23.4 and −14.1 dB, respectively, for excitation at P1, and −20.6 and −15.8 dB for excitation at P2. These results demonstrate a pseudospin-locked waveguide crossing that supports independent transmission through intersecting topological channels with strongly suppressed crosstalk.

## Conclusion

We have experimentally demonstrated pseudospin-locked chiral edge transport in gyromagnetic photonic crystals, combining pseudospin-selective excitation with nonreciprocal unidirectional propagation. Joint control of lattice deformation and magnetic bias governs dipolar–quadrupolar band inversions and provides access to the QSH, PQAH, and CI phases, enabling the design of interfaces with distinct pseudospin contributions. At the PQAH–CI interface, microwave near-field and transmission measurements reveal a single chiral edge branch that is selectively excited by a source with matching handedness. The persistence of edge transport in the presence of metallic defects further demonstrates robustness against the tested perturbations. By arranging domains with different spin Chern numbers, we extend this mechanism to a photonic pseudospin multiplexer, a complementary demultiplexer, and a waveguide crossing. These devices demonstrate that opposite pseudospin components can be combined into a common output channel, separated into distinct ports, and routed through intersecting interfaces while retaining their channel selectivity. The waveguide crossing exhibits average measured crosstalk levels of −14.1 and −15.8 dB for its two input channels, confirming independent transmission with suppressed interchannel coupling. These demonstrations connect pseudospin-resolved band engineering to experimentally realized routing functions and establish pseudospin as an addressable degree of freedom for organizing chiral photonic channels.

The distinction between chirality and pseudospin preservation remains central to this approach: the total Chern-number difference constrains the net chirality of the interface modes, whereas pseudospin selectivity additionally depends on the modal structure and sufficiently weak coupling between opposite pseudospin sectors. Extending the demonstrated devices to larger networks therefore calls for systematic characterization of pseudospin mixing at junctions, insertion loss, operating bandwidth, and accumulated crosstalk. Such studies would clarify how the observed interface-level robustness and channel selectivity translate into network performance. The present results provide an experimental foundation for nonreciprocal photonic networks that use pseudospin to control channel excitation, combination, separation, and routing.

**Acknowledgments**

Z.G. acknowledges funding from the National Key R&D Program of China (grant no. 2025YFA1412300), National Natural Science Foundation of China (grants no. 62361166627 and 62375118), Guangdong Basic and Applied Basic Research Foundation (grant no. 2024A1515012770), Shenzhen Science and Technology Innovation Commission (grants no. 202308073000209), and High-level Special Funds (grant no. G03034K004). X.X. acknowledges funding from the National Natural Science Foundation of China (grant no. 62405053), Guangdong Basic and Applied Basic Research Foundation (grant no. 2025A1515012229) and Youth S&T Talent Support Programme of Guangdong Provincial Association for Science and Technology (grant no. SKXRC2025440).

**Competing Interests**

The authors declare no competing interests.

## Main figure legends

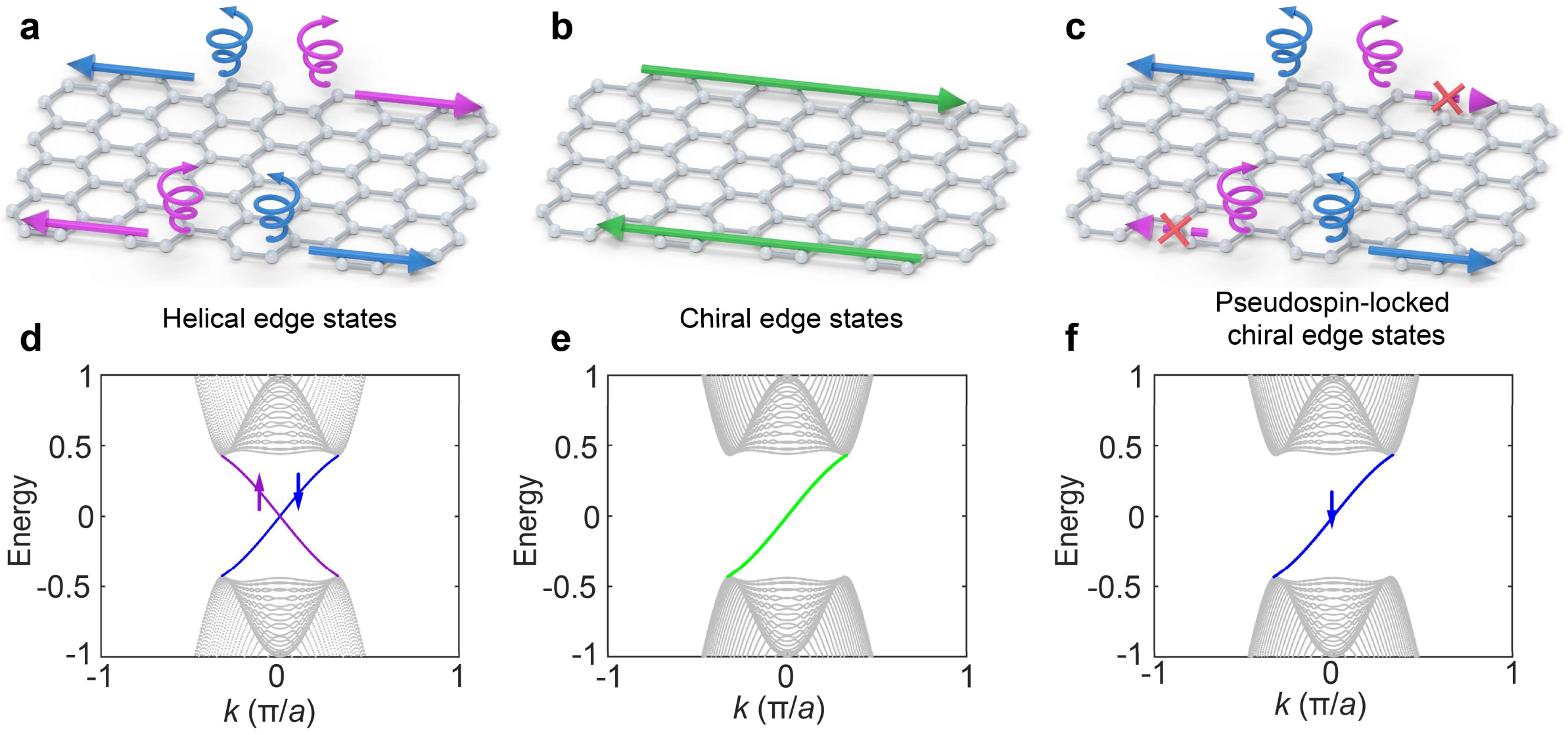

**Fig. 1 | Comparison of helical, chiral, and pseudospin-locked chiral edge states. a–c**, Schematics of helical (**a**), chiral (**b**), and pseudospin-locked chiral (**c**) edge states in photonic insulators. Magenta and blue helices represent pseudospin-up and pseudospin-down chiral sources, respectively. Straight arrows indicate propagation directions along the boundaries, and red crosses mark the absence of an edge channel for the indicated pseudospin. **d–f**, Corresponding representative projected band structures. Gray curves denote bulk bands, and colored curves denote edge modes. Magenta and blue branches represent pseudospin-up and pseudospin-down modes, respectively; the green branch represents the conventional chiral edge modes. The arrows on the magenta and blue branches indicate pseudospin labels.

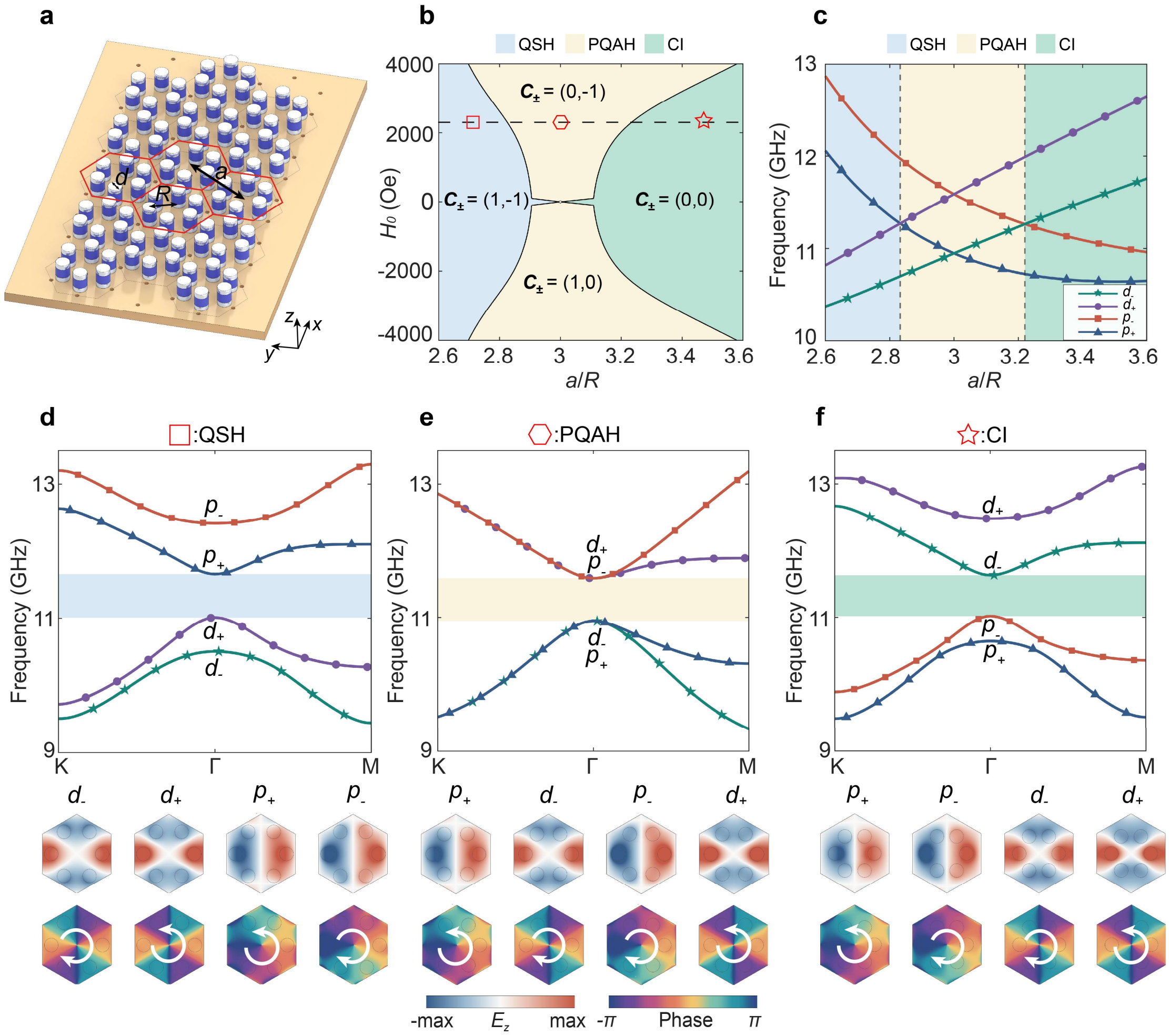

**Fig. 2 | Topological phase engineering in a gyromagnetic photonic crystal. a,** Schematic of the triangular-lattice gyromagnetic photonic crystal comprising hexagonal clusters of six yttrium iron garnet (YIG) rods. Each rod is sandwiched between two permanent magnets, and red hexagons outline the unit cells. The lattice constant is $a = 14$ mm, the rod diameter is $d = 2.8$ mm, and $R$ is the distance from each rod center to the unit-cell center. **b,** Phase diagram versus $a/R$ and the signed magnetic bias $H_0$, showing quantum spin Hall (QSH), pseudospin-polarized quantum anomalous Hall (PQAH), and conventional insulator (CI) phases. Their spin Chern numbers $(C_+, C_-)$ are $(1, -1)$ for QSH, $(0, -1)$ and $(1, 0)$ for PQAH under positive and negative bias, respectively, and $(0, 0)$ for CI. The horizontal dashed line marks $H_0 = 2200$ Oe; the red square, hexagon, and star identify the bulk band structures in **d**, **e**, and **f**, respectively. Bias reversal gives $(C_+, C_-) \rightarrow (-C_-, -C_+)$. **c,** Evolution of the four relevant Γ-point eigenfrequencies with $a/R$ at $H_0 = 2200$ Oe. Vertical dashed lines mark the gap-closing transitions between the QSH, PQAH, and CI regimes. **d–f,** Bulk band structures for $a/R = 2.70$ (QSH, **d**), 3.00 (PQAH, **e**), and 3.50 (CI, **f**), with $d/a = 0.2$. Shaded regions indicate the bandgaps. The corresponding Γ-point $E_z$ field profiles and phase distributions are shown below each band structure in ascending eigenfrequency. White curved arrows indicate the sense of phase winding.

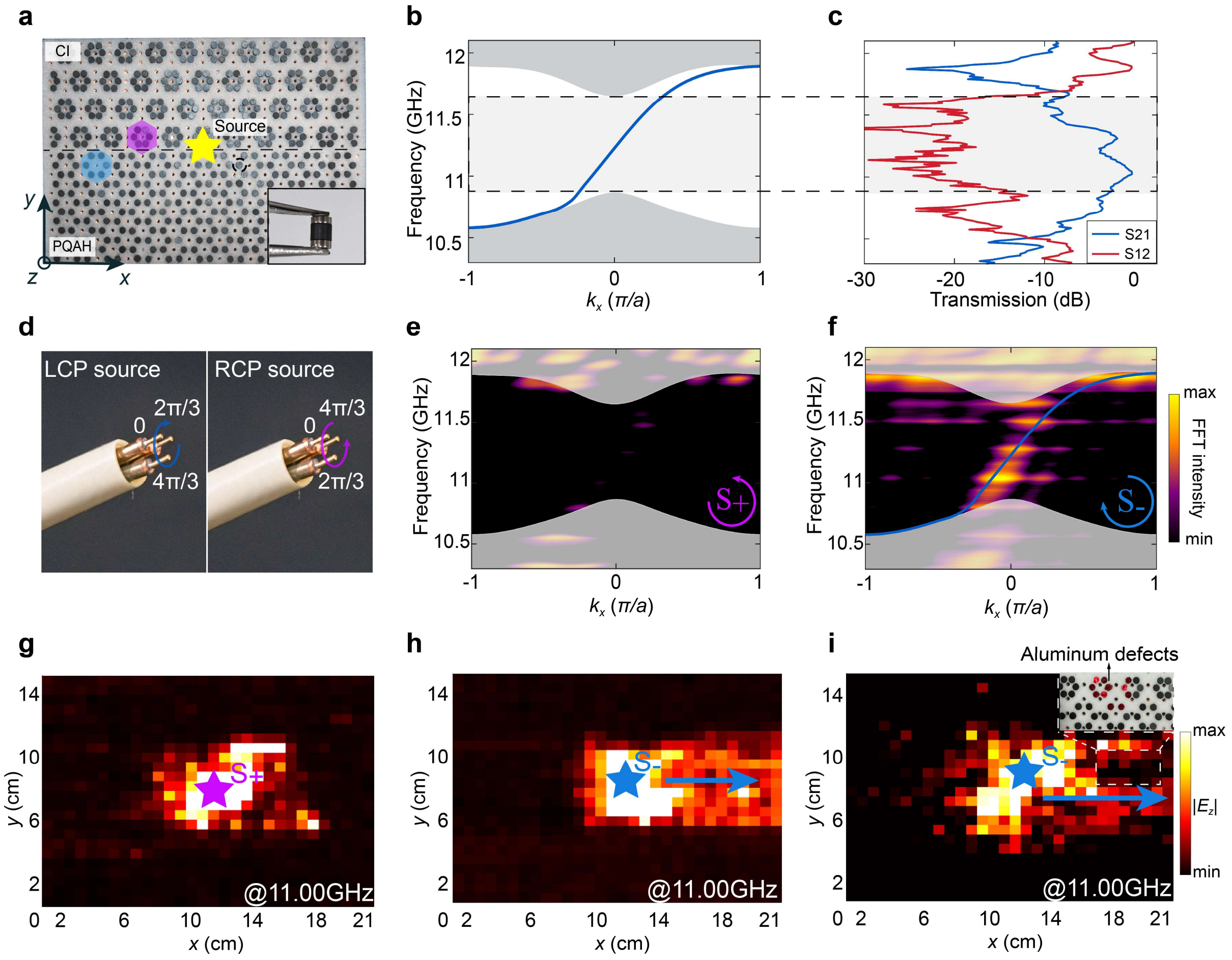

**Fig. 3 | Experimental observation of pseudospin-locked chiral edge states at a PQAH–CI interface. a,** Photograph of the sample, with the CI domain $(C_+, C_-) = (0, 0)$ above and the PQAH domain $(C_+, C_-) = (0, -1)$ below the dashed interface. Representative CI and PQAH unit cells are highlighted in purple and blue, respectively. The yellow star marks the source position, and the inset shows the rod–magnet assembly. **b,** Calculated projected band structure, with the chiral edge branch in blue and bulk bands in gray. **c,** Measured transmission spectra for rightward propagation, $S_{21}$ (blue), and leftward propagation, $S_{12}$ (red), plotted on the same vertical frequency scale as **b**. Horizontal dashed lines in **b,c** delimit the common bulk bandgap. **d,** Photographs of the three-element chiral sources with opposite phase sequences, labeled LCP and RCP according to the illustrated phase-circulation convention. **e,f,** Dispersion maps obtained by spatial Fourier transformation of the measured complex $E_z$ fields under pseudospin-up ($S_+$, **e**) and pseudospin-down ($S_-$, **f**) excitation. Gray shading indicates the calculated projected bulk bands, and the blue curve in **f** shows the calculated edge-state dispersion. The color scale represents Fourier-transform intensity. **g,h,** Measured field amplitudes $|E_z|$ at 11.00 GHz under $S_+$ and $S_-$ excitation, respectively. **i,** Measured $|E_z|$ at 11.00 GHz after selected boundary YIG rods are replaced by aluminum pillars, using the same $S_-$ excitation and source position as in **h**. The inset identifies the metallic defects. Stars mark source positions, and blue arrows indicate rightward propagation in **h,i**.

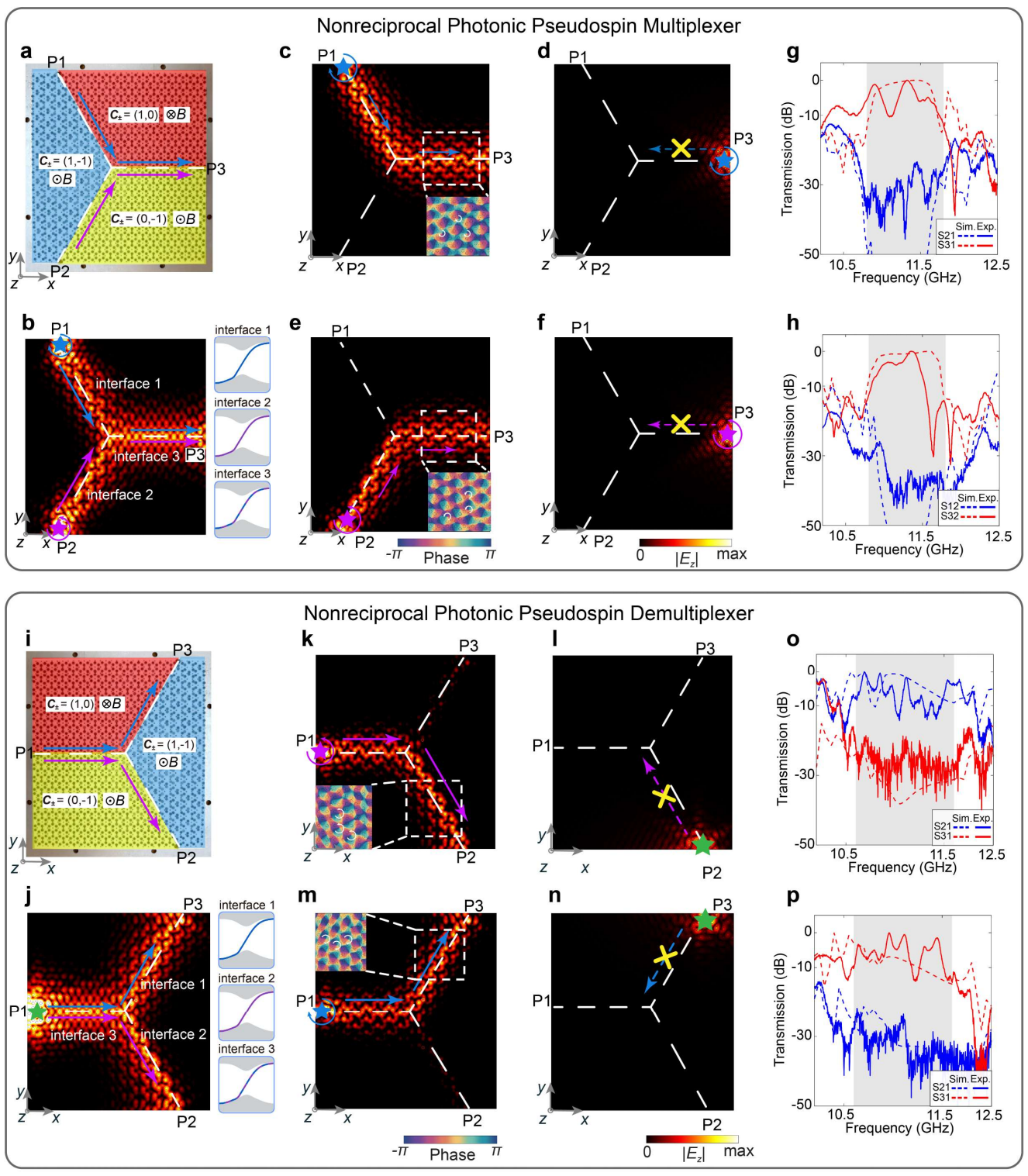

**Fig. 4 | Nonreciprocal photonic pseudospin multiplexing and demultiplexing at Y-junctions. a,i,** Annotated photographs of the multiplexer and demultiplexer, respectively, comprising three domains with spin Chern numbers $(C_+, C_-) = (1, -1)$ (blue), $(1, 0)$ (red), and $(0, -1)$ (yellow). The symbols $\odot B$ and $\otimes B$ indicate magnetic bias out of and into the plane, respectively. Magenta and blue arrows indicate pseudospin-up and pseudospin-down routing, respectively. **b,** Simulated field amplitude $|E_z|$ under simultaneous pseudospin-down excitation at P1 and pseudospin-up excitation at P2, showing transmission to the common output P3. The adjacent panels show the calculated dispersions for interfaces 1–3. **c,e,** Simulated $|E_z|$ under separate pseudospin-down excitation at P1 and pseudospin-up excitation at P2, respectively. Insets show the $E_z$ phase distributions within the dashed boxes in the output channel. **d,f,** Simulated $|E_z|$ under reverse excitation at P3 with pseudospin-down and pseudospin-up sources, respectively. Yellow crosses indicate suppressed propagation toward the junction. **g,h,** Simulated and measured transmission spectra for excitation at P1 and P2, respectively. Red curves show transmission to P3 ($S_{31}$ in **g** and $S_{32}$ in **h**), whereas blue curves show transmission to the other input port ($S_{21}$ in **g** and $S_{12}$ in **h**). **j,** Simulated $|E_z|$ showing separation of the two pseudospin components injected at P1 into outputs at P2 and P3. The adjacent panels show the calculated dispersions for interfaces 1–3. **k,m,** Simulated $|E_z|$ under pseudospin-up and pseudospin-down excitation at P1, showing routing to P2 and P3, respectively. Insets show the $E_z$ phase distributions within the dashed boxes. **l,n,** Simulated $|E_z|$ under reverse excitation at P2 and P3, respectively, showing suppressed propagation toward P1. **o,p,** Simulated and measured transmission spectra under pseudospin-up and pseudospin-down excitation at P1, respectively; blue and red curves denote $S_{21}$ and $S_{31}$. In **g,h,o,p**, dashed and solid curves represent simulations and measurements, respectively, and gray shading highlights the operating bands.

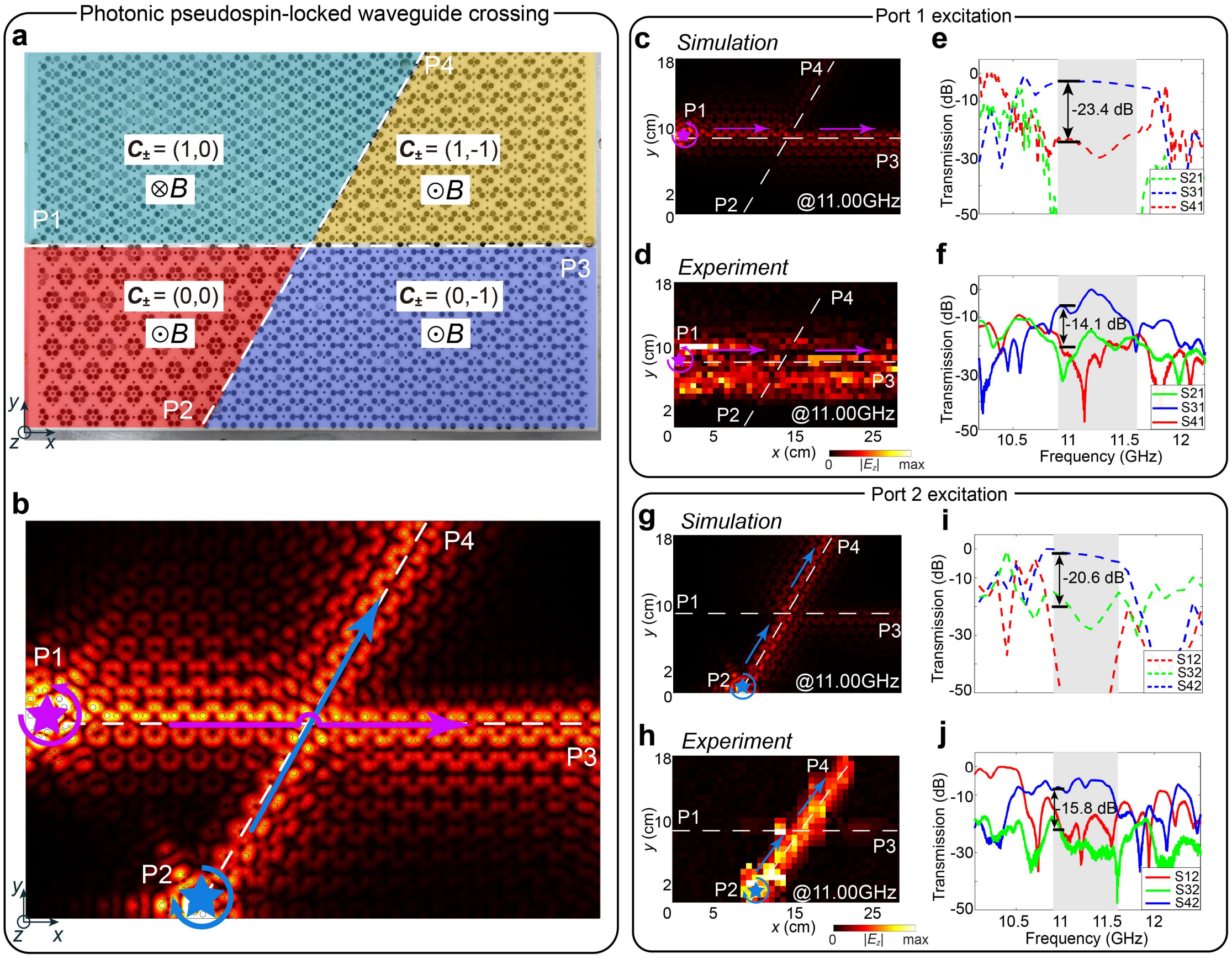

**Fig. 5 | Experimental demonstration of a pseudospin-locked photonic waveguide crossing.** **a,** Annotated photograph of the sample, comprising four domains with spin Chern numbers $(C_+, C_-) = (1,0)$, $(1,-1)$, $(0,0)$, and $(0,-1)$ in the upper-left, upper-right, lower-left, and lower-right regions, respectively. The symbols $\odot B$ and $\otimes B$ indicate magnetic bias out of and into the plane, respectively. The horizontal P1–P3 interface supports pseudospin-up propagation, whereas the intersecting P2–P4 interface supports pseudospin-down propagation. **b,** Simulated field amplitude $|E_z|$ under simultaneous pseudospin-up excitation at P1 and pseudospin-down excitation at P2, showing transmission through the crossing along the respective channels. Magenta and blue arrows indicate pseudospin-up and pseudospin-down propagation, respectively. **c,d,** Simulated and measured $|E_z|$, respectively, at 11.00 GHz under pseudospin-up excitation at P1. **e,f,** Corresponding simulated and measured transmission spectra, respectively; $S_{31}$ (blue) describes transmission to the intended output P3, whereas $S_{21}$ (green) and $S_{41}$ (red) describe transmission to the unintended ports. **g,h,** Simulated and measured $|E_z|$, respectively, at 11.00 GHz under pseudospin-down excitation at P2. **i,j,** Corresponding simulated and measured transmission spectra, respectively; $S_{42}$ (blue) describes transmission to the intended output P4, whereas $S_{12}$ (red) and $S_{32}$ (green) describe transmission to the unintended ports. Dashed curves in **e,i** are simulations, solid curves in **f,j** are measurements, and gray shading highlights the operating bands. The annotated crosstalk is −23.4 dB in simulation and −14.1 dB in experiment for P1 excitation, and −20.6 dB and −15.8 dB, respectively, for P2 excitation. Crosstalk is evaluated over 11.0–11.5 GHz by averaging each transmission spectrum in dB, taking the larger mean transmission to the two unintended ports, and subtracting the mean transmission to the intended output.